\documentclass[manuscript]{acmart}
\renewcommand\footnotetextcopyrightpermission[1]{}
\AtBeginDocument{%
  \providecommand\BibTeX{{%
    \normalfont B\kern-0.5em{\scshape i\kern-0.25em b}\kern-0.8em\TeX}}}

\usepackage{xspace}
\usepackage{mdframed}
\usepackage{color, colortbl}
\usepackage{array}
\usepackage{float}
\usepackage{enumitem}
\usepackage{graphicx}
\usepackage{multirow}
\usepackage{url}

\begin{document}

\mdfdefinestyle{MyFrame}{
 outerlinewidth=0pt,
 skipabove=0pt,
 skipbelow=0pt,
 innertopmargin=6pt,
 innerbottommargin=0pt,
 linewidth=0pt,
 topline=false,
 rightline=false,
 leftline=false,
 innerrightmargin=4pt,
 innerleftmargin=4pt}

\newcommand*{\tableIndent}{\hspace*{0.3cm}}
\newcommand{\GH}{{\sc GitHub}\xspace}
\newcommand{\MLH}{{\sc MLH}\xspace}
\newcommand{\DP}{{\sc Devpost}\xspace}
\newcounter{RQCounter}
\newcommand{\RQ}[2]{
\refstepcounter{RQCounter} \label{#1}
\begin{mdframed}[style=MyFrame]\noindent
	\textbf{RQ}$_{\arabic{RQCounter}}$.~\emph{#2}
\end{mdframed}
}
\newcommand{\rerq}[1]{\textbf{RQ}$_{\ref{#1}}$}
\newcounter{HCounter}
\newcommand{\Hyp}[2]{
\refstepcounter{HCounter} \label{#1}
\begin{mdframed}[style=MyFrame]\noindent
	\textbf{H}$_{\arabic{HCounter}}$.~\emph{#2}
\end{mdframed}
}

\newcommand{\rehyp}[1]{\textbf{H}$_{\ref{#1}}$}

\definecolor{Gray}{gray}{0.9}
\newcommand{\mysubsec}[1]{\smallskip \emph{\textbf{#1.}}}
\newcolumntype{L}[1]{>{\raggedright\let\newline\\\arraybackslash\hspace{0pt}}m{#1}}

\title{The CAT Effect: Exploring the Impact of Casual Affective Triggers on Online Surveys' Response Rates}


\author{Irene-Angelica Chounta}
\orcid{0000-0001-9159-0664}
\email{irene-angelica.chounta@uni-due.de}
\affiliation{%
  \institution{University of Duisburg-Essen}
  \city{Duisburg}
  \country{Germany}
}

\author{Alexander Nolte}
\orcid{0000-0003-1255-824X}
\email{alexander.nolte@ut.ee}
\affiliation{%
  \institution{University of Tartu}
  \city{Tartu}
  \country{Estonia}
}
\affiliation{%
  \institution{Carnegie Mellon University}
  \city{Pittsburgh}
  \state{PA}
  \country{USA}
}



\begin{abstract}
  We explore the impact of Casual Affective Triggers (CAT) on response rates of online surveys. As CAT, we refer to objects that can be included in survey participation invitations and trigger participants' affect. The hypothesis is that participants who receive CAT-enriched invitations are more likely to respond to a survey. We conducted a study where the control condition received invitations without affective triggers, and the experimental condition received CAT-enriched invitations. We differentiated the triggers within the experimental condition: one-third of the population received a personalized invitation, one-third received a picture of the surveyor's cat, and one-third received both. We followed up with a survey to validate our findings. Our results suggest that CATs have a positive impact on response rates. We did not find CATs to induce response bias.
\end{abstract}




\maketitle

\section{Introduction}
\label{sec:intro}
Surveys are a standard and informative tool for gathering feedback, measuring satisfaction, and gaining insight regarding the intentions of individuals in various domains, such as Human-Computer Interaction (HCI)~\cite{muller2015}, marketing~\cite{evans2018value}, education~\cite{chounta2021exploring} and others. Online surveys, in particular, are popular because they are cost-efficient, they can be used to reach out to wider audiences regardless of spatial and temporal restrictions, and the distribution and data-collection processes are straightforward and effective~\cite{wright2005researching}. Research on surveys typically focuses on survey design in terms of content. This includes what questions to ask, in what way, and in what turn to retrieve reliable and informative feedback from the target audience~\cite{mueller2014}. It also includes aesthetic and presentation-related aspects of survey design~\cite{guin2012myths}. However, there is also research that focuses on studying response rates and means of enticing them. 

Response rates have long been utilized as a quality indicator for surveys~\cite{biemer2003introduction}. Newer works indicate that response rates might not have as strong an effect as commonly expected~\cite{groves2006nonresponse}, but declining response rates have nonetheless become a concern~\cite{de2002trends}. This led to researchers developing and studying a variety of means to entice responses. Common means of improving response rates are monetary incentives or rewards~\cite{guo2016population} and targeted email invitations. For example, Petrov\v{c}i\v{c} et al.~\cite{petrovvcivc2016effect} studied how survey invitation design can affect response rates when using textual elements that convey authority, a sense of community, or a plea for help. According to their findings, pleading for help increased the response rates but using more than one element did not necessarily improve response rates further. Moreover, there are studies that report no significant effect of incentives on response rates in the context of organizational research~\cite{baruch2008survey}. Findings related to the effects of different incentives are thus still inconclusive.

Adding to this area of research, we developed the concept of casual affective triggers (CATs). Utilizing different CATs, we aim to improve online survey response rates in terms of partial and complete survey participation. As CAT, we define informal triggers such as the display of simple (or else, 
\textit{casual}) objects or artifacts that can be included in survey email invitations to trigger participants' affect; for example, text or images that induce emotional responses. The CAT concept is based on prior work in the contexts of collaborative learning~\cite{pekrun2002academic}, environmental~\cite{han2017cognitive}, and social psychology~\cite{perugini2001role} which points towards the influence of positive and negative affect on behavior and behavioral intentions of individuals. The rationale is that affect plays an important role regarding participants' immediate or direct reactions -- in this case, survey response. Thus, the hypothesis is that participants who receive CAT-enhanced survey invitations (that is, invitations containing some kind of casual affective trigger) are more likely to respond to a survey than those who receive invitations with no CAT element. 
 
For this research, we designed a $2x2$ factorial study to explore the impact of two CATs and their combined effect on online survey response rates. The one trigger (CAT1) aimed at positive feelings conveyed by personalization~\cite{joinson2007personalized}, and the second trigger (CAT2) aimed at positive feelings that could be prompted by the picture of the surveyor's cat~\cite{oliver2011entertainment}. Then, we designed four different types of email invitations: 1) without any CAT, 2) a personalized email invitation (CAT1), 3) an email invitation that contained a picture of the surveyor's cat as an attachment and a related greeting (CAT2), 4) a personalized email invitation that contained a picture of the surveyor's cat as an attachment along with a related greeting (CAT1 and CAT2). With this study, we thus aim to answer the following research questions:

\RQ{RQ1}{Do CATs have a significant impact on online survey response rates?}
\RQ{RQ2}{Do CATs have an additive effect on online survey response rates when combined?}

Using CATs might, however, affect individuals' response behavior. They might, in particular, induce response bias~\cite{furnham1986response} in that individuals that receive CATs might respond favorably to survey questions compared to those that did not receive such triggers. To assess whether CATs indeed induce response bias, we also aim to answer the following third research question:

\RQ{RQ3}{Do CATs introduce bias in online survey responses?}

This work is based on two assumptions: (a) that there is a link between the proposed CATs and positive emotions and (b) that there is a link between positive emotions and survey responses.   

Regarding the first assumption (a), we built on prior research that provides evidence for the link between the CATs used in this work and positive emotions. On the one hand, personalized salutations (i.e., calling one by their name) can convey positive emotions, which can lead to participants feeling more important and valued or more “special”~\cite{joinson2007personalized}. On the other hand, Myrick~\cite{myrick2015emotion} demonstrated that viewing cat-related media online has a positive influence on affect (that is, it triggers enjoyment and positive emotions). Their work shows that people who own or used to own a cat are more likely to view cat-related media. At the same time, they do not report any negative effects on people who do not like cats. 

Regarding the second assumption (b), we used as a basis existing research that suggests that positive affect broadens attention and cognition reinforcing creativity and flexibility, as well as thought-action urges~\cite{fredrickson2005positive}. We argue that in our context, thought-action urges may translate into survey participation.

To support these assumptions, validate our findings and further clarify the potential relationships between survey design principles, casual affective triggers and response rates, we followed up with a second online survey distributed using a crowdsourcing platform.

The remainder of the paper is structured as follows: in the next section, we provide an overview of related work on online surveys (section~\ref{sec:background}). Next, we present the methodology and experimental setup of our study (section~\ref{sec:method}), and we follow up with our findings (section~\ref{sec:results}). We provide a contextualized discussion on the findings including a discussion of the contribution and limitations of this work as well as its theoretical and practical implications (section~\ref{sec:discussion}). The paper ends with concluding remarks (section~\ref{sec:conclusion}).

\section{Related Work}
\label{sec:background}
Identifying factors and aspects that can encourage individuals to participate in surveys has been a topic of interest for multiple decades~\cite{fox1988mail,schuldt1994electronic,fan2010factors,petrovvcivc2016effect}. Starting from mail~\cite{schuldt1994electronic} and telephone surveys~\cite{curtin2005changes} to the nowadays prevalent online surveys~\cite{fox1988mail,wright2005researching}, researchers have studied how various aspects of survey design can affect response rates. We perceive response rate as the number of completed surveys divided by the number of surveys that were sent and not returned as undeliverable (adapted from~\cite{kaplowitz2004comparison}). Scholars report various levels of survey responses in different contexts ranging from less than 1\%~\cite{koo2005challenges} in the medical domain and 5\%~\cite{de2007satisfaction} in management to 9\%~\cite{zillmann2014survey} for a survey on online dating, 10\%~\cite{qiu2019going} in the context of open source communities, 18\%~\cite{qiu2021approaches} for a survey on coding events and sometimes even more than 60\%~\cite{jeppesen2006users} in organizational science.

Related work reports on four categories of factors related to response rates. These categories cover survey development, delivery, completion, and return~\cite{fan2010factors}. The delivery aspect can further be divided into sampling (e.g., based on age~\cite{kaldenberg1994mail}), delivery (e.g., via mail, phone, or email~\cite{yammarino1991understanding}), invitation design, pre-notifications and reminders, and incentives~\cite{fan2010factors}. Our work particularly focuses on invitation design.

Related to invitation design, researchers studied the effect of various aspects on response rates. These include technical aspects, such as providing a realistic time estimate to complete the survey, an explanation for where researchers obtained email addresses and providing the researcher contact information on survey response rates~\cite{heerwegh2005effects,crawford2001web}. They also include aspects related to utilizing empty email subject lines to cause curiosity ~\cite{porter2005mail} and hinting towards the scarcity of the selected group~\cite{porter2003impact}. However, prior work did not intensively study the potential effect of affective triggers~\cite{perugini2001role,han2017cognitive} on response rates.

Related research discusses the potential impact of personalization of invitation emails on survey response rates. The reported findings are ambiguous in that there are studies that did not find an effect of personalization on response rates~\cite{trespalacios2016effects}. Some studies indicate improved response rates when invitations are personalized, especially when researcher(s) and participants have prior relations~\cite{heerwegh2005effects}. Our aim is to add to these findings from the perspective of affective triggers.

Since the emergence of online surveys, researchers have studied how to utilize images~\cite{witte2004instrument}. Guidelines such as the ones formulated by Kaczmirek~\cite{kaczmirek2005web} often suggest abstaining from sending invitations that contain graphical content because of -- among other reasons -- their potential to induce bias. Related studies indeed show that images included in the main body of the survey can induce a bias when the image is associated with the content of the survey~\cite{couper2007visual}. However, other reports indicate that including general or abstract images in survey invitations can foster participation compared to concrete images~\cite{liu2016impact}. We aim to contribute to this line of work by exploring if specific graphical content can serve as an incentive for participation without inducing bias.

The design of a survey and survey invitation can induce several biases, which, in turn, can affect the usefulness of the findings derived from the survey~\cite{deming1944errors}. The two most common biases discussed in related work on the design of invitations are non-response bias and response bias. Non-response bias can occur when individuals who do not participate in a survey are systematically different from those that participate~\cite{sheikh1981investigating}. In comparison, response bias occurs when participants respond inaccurately or falsely to survey questions~\cite{furnham1986response}. The latter is particularly important in the context we study because affective triggers are designed to cause an emotional response. This emotional response can influence how individuals answer to survey questions, as has been found when studying the effect of incentives on response bias~\cite{mizes1984incentives}.

\section{Methodology}
\label{sec:method}
For our study, we focused on hackathon participants (sampling) who received an email invitation (delivery) with no additional incentives to participate. We designed the email invitation according to guidelines proposed in prior work with the aim to maximize participation. In particular: a) we indicated the scarcity of the survey in that we deliberately mentioned that only participants of a particular hackathon received an invitation \cite{porter2003impact}; b) we indicated where we obtained the email addresses; c) we provided an estimated duration for completing the survey; and d) we provided contact information \cite{crawford2001web,heerwegh2005effects}. We chose these recommendations because they were simple and easy to implement. 

To explore our goal, we formulated the following hypotheses:
\Hyp{H1}{The use of a personalized email invitation (CAT1) will have a positive impact on response rates in terms of partial and complete survey participation.}
\Hyp{H2}{The use of a cat picture (CAT2) will have a positive impact on response rates in terms of partial and complete survey participation.}
\Hyp{H3}{The combined use of a personalized email (CAT1) and cat picture (CAT2) will have an additive positive impact on response rates in terms of partial and complete survey participation.}
\Hyp{H4}{The survey responses in terms of item ratings will significantly differ among the control and experimental conditions.}

\subsection{Online Survey}
We invited 4502 participants to respond to an online survey to investigate the research hypotheses. The participants were selected based on their previous participation in two separate online hackathons (Hack1 and Hack2). Hackathons have become popular time-bounded events during which participants form teams and engage in intense collaboration over a short period to complete a project of interest to them~\cite{nolte2020organize}. These projects often evolve around developing some for of software prototype~\cite{nolte2020happens}. The hackathons we studied were organized by the same organization. They took place at the beginning of the COVID-19 crisis -- at the end of March 2020 (Hack1) and at the beginning of April 2020 (Hack2) respectively -- and were mainly attended by students and young entrepreneurs. Hack1 had more than 4000 registered participants and Hack2 more than 400. Both hackathons took place online so participants could theoretically join from any location if they had a sufficiently stable internet connection. Still, the events had a local focus in that they were co-organized by organizations that aim to develop the entrepreneurial ecosystem within a specific geographic region. Consequently, the majority of participants came from the respective region. The hackathons were competitive in that participating teams could win monetary prizes. A jury chose the winners based on video presentations that participants submitted at the end of the event.

The online survey aimed to assess the participants' experience during the hackathon. For this purpose, we adapted established multi-point Likert scales, which were anchored between "strongly disagree" (1) and "strongly agree" (5). We particularly included scales related to the participants' motivation to participate in the hackathon~\cite{filippova2017diversity,clary1992volunteers}, their perception of the team process~\cite{filippova2017diversity}, their perception about their influence in team decisions~\cite{filippova2017diversity}, their satisfaction with the outcome they achieved~\cite{filippova2017diversity}, their intentions to continue working on the project they had started during the hackathon after the event had ended~\cite{bhattacherjee2001understanding} and their perception about the fairness of the judging process~\cite{landy1978correlates,leventhal1980beyond,greenberg1986determinants}. We also asked participants about their general level of anxiety related to the global pandemic\footnote{https://www.psychiatry.org/newsroom/news-releases/new-poll-covid-19-impacting-mental-well-being-americans-feeling-anxious-especially-for-loved-ones-older-adults-are-less-anxious} and about their feeling of community related to the other participants of the hackathon~\cite{ellison2007benefits}.

\subsection{Study Design}
For our research purposes, we designed a 2x2 factorial experiment to explore two types of affective triggers: 
\begin{itemize}
    \item a personalized invitation (CAT1). We personalized the invitation to the survey by using the participant's first name as a greeting. By adding this, we aimed to create a positive emotion to the participant due to individual and targeted treatment. Personalized invitations have been used in marketing for increasing marketing metrics~\cite{heerwegh2005effects};
    \item a cat picture (CAT2). We attached in the email invitation a picture of the surveyor's cat along with a reference in the email. The reference read, \textit{``Me and my cat are looking forward to your answers and thank you for your time and help!"} while in emails that did not contain a cat picture, the reference was \textit{``I look forward to your answers and thank you for your time and help!"}. By adding the cat picture, we aimed to create a positive emotion to the participant.
\end{itemize} 

\begin{figure}[htb]
\centerline{\includegraphics[width=\linewidth,keepaspectratio,]{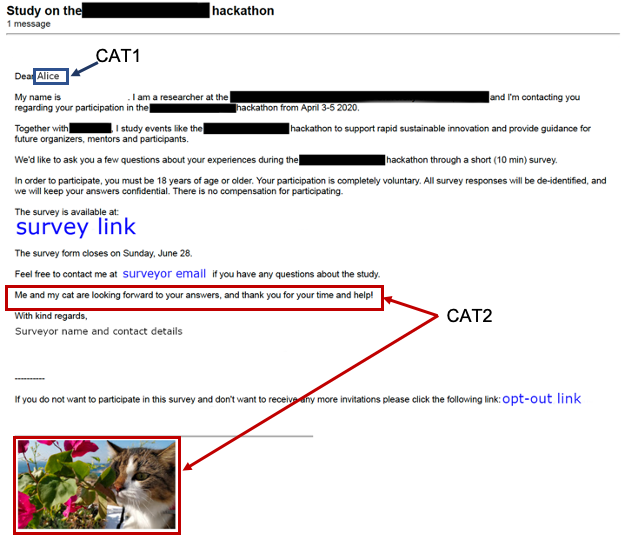}}
\caption{Example of email invitation with all CAT elements.}
\label{fig:cats}
\Description{A screenshot of the email invitation that shows both affective triggers used in the study. The email starts with a personal salutation to the email recipient, followed by the email body that provides information about the survey and the surveyor, and invites the recipient to participate following a web link. At the bottom, the picture of a cat is displayed.}
\end{figure}

Additionally, we explored the combination of a personalized invitation (CAT1) and a cat picture (CAT2) in order to study further potential additive effects from the combined use of different kinds of affective triggers. An example invitation with all the CATs used is presented in Figure \ref{fig:cats}.

Based on the 2x2 factorial design, we split the study population into two conditions:
\begin{itemize}
    \item Control Condition: The participants in the control condition (922 individuals) received no CAT-enhanced email. Instead, they received a non-personalized invitation with no picture attached.
    \item Experimental Condition: The participants in the experimental condition (3580 individuals) received either a personalized invitation (CAT1, 1330 individuals), an invitation accompanied by a cat picture (CAT2, 922 individuals), or a personalized invitation that was accompanied by a cat picture (CAT1 and CAT2, 1328 individuals).
\end{itemize}

From the initial set of invitations sent, 269 email invitations bounced back. After removing the participants whose emails bounced back, the dataset consisted of 4233 participants over the two conditions. The overview of the study design and setup is presented in Table \ref{tab:study}. We followed up with a reminder 14 days after the first invitation for the participants who had not yet responded (that is, 4130 reminders). The reminder followed the same approach as the original invitation with respect to the conditions -- that is, the same people received the same triggers as in the first invitation. To obtain consent, we followed the US Code of Federal regulations~\footnote{45 CFR §46.116(c) General requirements for informed consent.} by informing participants about the dual nature of our study through a debrief email.

\begin{table*}[h]
\begin{center}
\begin{tabular}{L{3cm}ccc}
\toprule 
     & \multicolumn{1}{c}{\textbf{CAT1 = 0}} & \multicolumn{1}{c}{\textbf{CAT1 = 1}} \\
    
   \textbf{CAT2 = 0}& N = 875 & N = 1234   \\
    \textbf{CAT2 = 1} & N = 870 & N = 1254  \\
   
    \bottomrule
    
\end{tabular}
\end{center}
\caption{The 2x2 factorial study design that was implemented in this research}
\label{tab:study}
\end{table*}

We distinguished two cases regarding participants' response behavior: (a) participants who accessed the survey but did not necessarily respond to all its sections. From now on, we will refer to this case as: \textit{``Partial Participation"}, and (b) participants who responded to all the sections of the survey. From now on, we will refer to this case as: \textit{``Complete Participation"}. We analyzed these two cases separately in order to gain insights regarding short-term effects on participants that do not lead to successful survey completion.

\subsection{Analysis}
To investigate the first three research hypotheses (\rehyp{H1}, \rehyp{H2}, \rehyp{H3}, section \ref{sec:background}), we carried out an analysis of the descriptive statistics regarding the response rates of the survey per condition (control and experimental) and case (partial and complete participation). Then, we performed regression analysis to assess the impact of affective triggers on survey response rates. 
For all participants, we used the regression model to analyze the independent effects of CAT1 and CAT2. Then, specifically for investigating \rehyp{H3}, we introduced an interaction term to the initial regression model to test for any additive effect between CAT1 and CAT2. 

To investigate the final research hypothesis \rehyp{H4} (section \ref{sec:background}), we conducted a one-way ANOVA for the participants who completed the survey. To that end, we calculated the average score for each survey category (for example, motivation, perception of the team process, and so on) and the overall average score for all the survey items. Then we compared the mean scores within groups. The null hypothesis (\textbf{$H_0$}) of the ANOVA is that there is no difference in means. The alternate hypothesis (\textbf{$H_a$}) is that the means are different thus the presence of affective triggers introduces bias in terms of survey responses.

\subsection{Validation}
In the main survey, we intentionally abstained from including questions that were unrelated to the survey topic to avoid response bias (\rerq{RQ3}). However, this decision meant that it was necessary to further clarify the potential relationships between survey design elements, affective triggers, and response rates. In particular, we needed to clarify:
\begin{enumerate}[label=C\arabic*]
    \item The connection between different triggers we utilized and individual participation intentions. With the main survey we study how CAT1 (personalized invitation), CAT2 (a cat picture) and their combination affect survey participation. The invitation also included additional triggers that have been found to increase response rates though (c.f. section \ref{sec:background}). We thus needed to establish whether and how the different triggers might affect individual participation intentions. The respective triggers are:
    \begin{itemize}
        \item Personally addressing the invited individual (CAT1)
        \item Attaching the picture of a cat (CAT2)
        \item Hinting towards the scarcity of the invited group of individuals
        \item Providing an explanation from where the utilized contact information was obtained
        \item Providing a realistic time estimate
        \item Providing contact information of the surveyor
    \end{itemize}
    From the main survey, we would not be able to understand the individual effects of the different utilized triggers on participation intentions. Instead, we would only see whether someone participated or not, but we would not know why.
    \item The relationship between individual preferences for pictures of cats and their perception of the picture as a design element for the survey invitation. This is necessary because individuals might have particular preferences for pictures of dogs and might in turn not appreciate seeing the picture of a cat. Since the main survey only included the picture of a cat, we would not be able to observe this potential relationship.
    \item The relationship between individual participation intentions and the preferences of individuals for pictures of cats, dogs, and small animals. From the main survey, we would not be able to make this distinction due to the same reason as mentioned before: we only utilized the picture of a cat.
\end{enumerate}

\begin{figure}[htb]
\centerline{\includegraphics[width=\linewidth,keepaspectratio,]{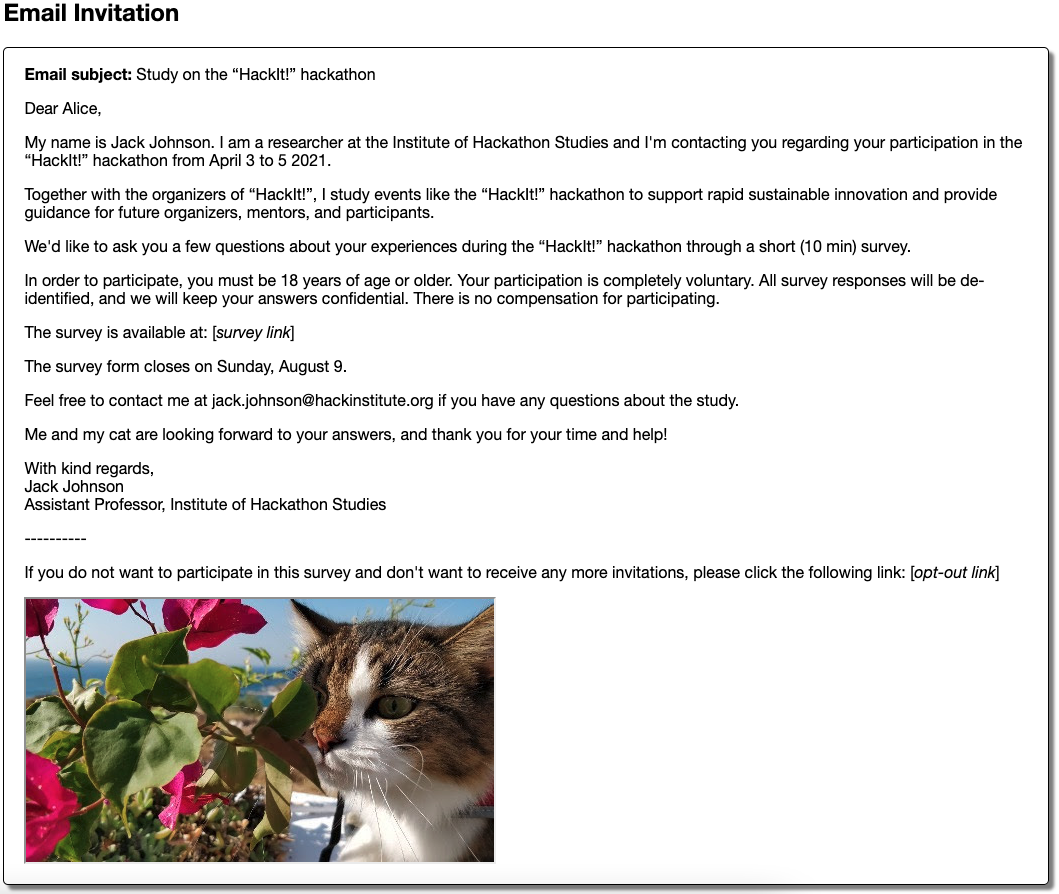}}
\caption{Example of email invitation demonstrated during the validation study with all CAT elements.}
\label{fig:survey2_invitation}
\Description{A screenshot of the email invitation that was used during the validation study. The email starts with a personal salutation to the email recipient, followed by the email body that provides information about the survey and the surveyor and invites the recipient to participate following a web link. At the bottom, the picture of a cat is displayed.}
\end{figure}

To clarify these relationships, we designed a second survey (c.f. Table~\ref{tab:app:instrument} in appendix~\ref{sec:app:instrument} for an overview of the questions included) which we distributed via a popular crowdsourcing platform. For participation, we provided monetary compensation. The second survey contained the same invitation text and picture that we utilized in the main survey (c.f. Fig~\ref{fig:survey2_invitation}). Instead of referring to a real hackathon, we referred to a fake event and asked the participants to imagine that they participated in this event and that they would receive the displayed survey invitation afterwards.

In addition to typical demographics, we included three separate questions to clarify the previously discussed relationships. First, we asked participants about their intentions to participate in a survey after receiving an invitation like the one displayed in Fig.~\ref{fig:survey2_invitation}. As answer options, we utilized a Likert scale which was anchored between "strongly disagree" (1) and "strongly agree" (5). In addition, we included a follow-up question where we asked them specifically about which of the design aspects of the survey would mainly lead them to participate in the survey they received an invitation for. In particular, we asked them about the:
\begin{itemize}
    \item Personalized salutation (CAT1)
    \item Attached picture (CAT2)
    \item Targeted audience scarcity
    \item Source of information
    \item Realistic time estimate
    \item Investigator information
\end{itemize}
As answer options, we utilized the same Likert 5-point scale anchored between "strongly disagree" (1) and "strongly agree" (5).

Finally, we included a scale asking participants about their feelings towards pictures of cats, dogs, and small animals. For this we utilized a common hedonic scale anchored between "dislike extremely" (1) and "like extremely" (9)~\cite{peryam1957hedonic}. We included these particular examples because prior work indicates that people who enjoy viewing cat-related media also enjoy watching videos or viewing pictures of dogs or other (small) animals~\cite{myrick2015emotion}. We expect to find the same similarities in the context of our study.

We also included quality control, attention-check questions and removed user responses that failed this quality control \cite{berinsky2014separating}: for example, we included an item that asked participants to provide a "strongly disagree" rating and we removed the responses that did not provide the requested rating. Following this process, we collected a total of 100 responses. To ensure that the order of the survey items would not affect the participants' responses, we split the population into two
groups. The first group (50 participants) received the questions in the order \textit{response intention, design aspects, pictures of small animals and affect, demographics} while the second group (50 participants) received the questions in the order \textit{pictures of small animals and affect, response intentions, design aspects, demographics}. The section \textit{demographics} was presented last in both groups to minimize cognitive load and fatigue.

After calculating common descriptives, we carried out three separate regression analyses to clarify the three different aspects: C1, C2, and C3. With the first we aimed to investigate the impact of survey design elements on participation intentions (C1). For this we utilized ordinal logistic regression since the dependent variable (\textit{Participation Intentions}) is ordinal~\cite{mccullagh1980regression}. The second was split into two parts. We (1) analyzed the relationships between the preference of participants for pictures of cats, dogs, and small animals, and the rating of the cat picture as a design element for the survey invitation (C2), and (2) potential relationships between the preference of participants for pictures of cats, dogs, and small animals and participation intentions (C3). For both, we utilized ordinal logistic regression models.

\section{Results}
\label{sec:results}
Here, we present the results of the experimental study. First, we present the results for partial participation -- that is, we focus on participants who accessed the survey but did not complete it (section~\ref{sec:results:partial}). Then, we present the results for complete participation -- that is, we focus on participants who completed the survey (section~\ref{sec:results:complete}). In both cases (partial and complete participation), our findings suggest that the use of CATs had a positive, statistically significant effect on response rates. The use of CAT1 appeared to have a strong, immediate effect that wears off over time, while the effect of CAT2 appeared to be constant over time. Further analysis indicated that this positive effect did not introduce undesirable bias in survey responses. We report the results of the analysis after adjusting the significance levels according to the Benjamini \& Hochberg (BH) correction for multiple comparisons~\cite{benjamini1995controlling}. A follow-up validation study confirmed our findings (section~\ref{sec:results:validation}).

Overall, only 5\% of the participants -- that is, 229 out of 4233 people -- accessed the survey at least one time. This rate is similar albeit lower than the ones reported by studies in similar contexts such as coding events (18\%,~\cite{qiu2021approaches}) and open source communities (10.7\%,~\cite{qiu2019going}). Our participant demographics included 20.48\% female and 76.51\% male participants, with the remaining 3.01\% abstaining from identifying their gender. Most participants reported being between 18 and 24 (51.81\%) and 25 and 34 years old (27.11\%), with fewer participants reporting to be between 35 and 44 (11.45\%), 45 and 54 (4.22\%) and 55 and 64 (3.61\%) years old. Few participants abstained from identifying their age (1.80\%).

\subsection{Partial Participation}
\label{sec:results:partial}
From the 229 participants that accessed the survey at least one time, 29 out of 229 (13\%) were in the control condition, 65 participants (28\%) received a personalized invitation (CAT1), 37 participants (16\%) received an invitation accompanied by a cat picture (CAT2) and 98 participants (43\%) received a personalized invitation along with a cat picture (CAT1 and CAT2). The analysis of the descriptive statistics suggests that the response rates for the experimental condition are higher than the control condition (Table \ref{tab:partialresponses} provides an overview). Participants who received CAT-enhanced invitations accessed the survey more than participants who did not receive any affective triggers. Participants who received multiple affective triggers (CAT1 and CAT2) have the highest response rate compared to the other participants. 
\begin{table*}[h]
\begin{center}
\begin{tabular}{L{2.5cm}lrrrlrr}
\toprule 
 & \multicolumn{1}{c}{\textbf{Control Condition}} & \multicolumn{3}{c}{\textbf{Experimental Condition}} & \multicolumn{3}{c}{\textbf{All Responses}}\\
    \addlinespace[0.1cm]
   & & \multicolumn{1}{l}{CAT1} & \multicolumn{1}{l}{CAT2} & \multicolumn{1}{l}{CAT1 and CAT2} \\
    Before Reminders        &  17 (1.94\%) & 48 (3.89\%) & 26 (2.99\%) & 60 (4.78\%) & 151 (3.57\%)    \\
    After Reminders &  12 (1.39\%) & 17 (1.43\%) & 11 (1.3\%) & 38 (3.18\%) & 78 (1.91\%)   \\
    Total Responses  &  29 (3.31\%)& 65 (5.27\%)& 37 (4.25\%) & 98 (7.81\%) & 229 (5.41\%)  \\

    \bottomrule
    
\end{tabular}
\end{center}
\caption{Number of partial survey responses per condition before and after the reminder that was sent 14 days after the initial invitation. In parentheses, we provide the response rate per condition, that is, the percentage of partial responses per number of participants for each design group.}
\label{tab:partialresponses}
\end{table*}

\subsubsection{CAT Impact on Partial Participation}
We used logistic regression to classify participants regarding their response behavior. That is, we defined two classes of response behavior:
\begin{itemize}
    \item Class 0: Participants who did not access the survey at all;
    \item Class 1: Participants who accessed, at least, the first section of the survey.
\end{itemize}

We trained a binary classifier to assess the impact of the two affective triggers examined in this study, namely CAT1 and CAT2. We modeled the email reminders as a random effect since reminders were not constant for each individual -- that is, those participants who did not complete or did not respond to the survey after the first invitation received email reminders.

\begin{table*}[h]
\begin{center}
\begin{tabular}{lcccccc}
\toprule 
\multicolumn{1}{c}{\textbf{Random Effects}}& \multicolumn{3}{c}{\textbf{Partial Participation}} & \multicolumn{3}{c}{\textbf{Complete Participation}} \\
    \addlinespace[0.1cm]
    & \multicolumn{1}{c}{Variance} & \multicolumn{1}{c}{Std. Dev} & \multicolumn{1}{c}{} & \multicolumn{1}{c}{Variance} & \multicolumn{1}{c}{Std. Dev} & \multicolumn{1}{c}{}\\
Reminder(Intercept)  & 3.001  &  1.732 &  &  3.375  &1.837  &   \\

   \multicolumn{1}{c}{\textbf{Fixed Effects}}& \multicolumn{3}{c}{} & \multicolumn{3}{c}{} \\
    \addlinespace[0.1cm]
    & \multicolumn{1}{c}{Coeffs} & \multicolumn{1}{c}{Std. Error} & \multicolumn{1}{c}{z-value} & \multicolumn{1}{c}{Coeffs} & \multicolumn{1}{c}{Std. Error} & \multicolumn{1}{c}{z-value} \\
    
    Intercept  & -1.93  & 1.31 & -1.81 &-2.23 $^{.}$ & 1.23 & -1.47\\
    CAT1  & 0.57 $^{*}$ & 0.16 & 1.98 &0.38 $^{*}$ & 0.18 &  2.34 \\
    CAT2      & 0.35  $^{*}$ & 0.15 & 2.13 &0.34 $^{*}$  & 0.17& 3.51    \\
  
    \bottomrule
    \multicolumn{5}{l}{\scriptsize{$^{***}p<0.001$, $^{**}p<0.01$, $^*p<0.05$, $^.p<0.1$}}
\end{tabular}
\end{center}
\caption{Regression models for partial participation and complete participation that explored the impact of affective triggers (CAT1 and CAT2) on response behavior. Email reminders were modeled as a random effect.}
\label{tab:partialregr}
\end{table*}

The results of the regression analysis on partial participation (Table \ref{tab:partialregr}) indicate that personalized invitations (CAT1) had a statistically significant positive effect on partial participation. In particular, the odds of responding to a personalized invitation were 77\% higher than the odds of responding to a non-personalized invitation. 
Invitations accompanied by the cat picture (CAT2) also had a statistically significant positive effect on responses; the odds of responding to an invitation accompanied by a cat picture were 42\% higher than the odds of responding to an invitation without a cat picture.

To explore the potential interaction between CAT1 and CAT2, we introduced an interaction term ($CAT1*CAT2$) to the logistic regression model~\cite{jaccard2001interaction}. Here, an interaction would occur if the relation between one predictor (CAT1) and the outcome variable (survey response) depended on the value of the other independent variable (CAT2). The results suggested that the interaction term was not statistically significant. Finally, we compared the two models (with and without the interaction term) using ANOVA. The results suggested that the two models were not significantly different, and therefore, we can ignore the interaction term $CAT1*CAT2$. The results of the analysis with the interaction term are presented in Table \ref{tab:interact} and Appendix \ref{sec:app:interaction}, Table \ref{tab:anovamodels}.

\subsection{Complete Participation}
\label{sec:results:complete}
Overall, only 4\% of the participants (that is, 171 out of 4233 people) completed the survey. 23 out of the 171 participants that completed the survey (13\%) were in the control condition, 47 participants (27\%) received a personalized invitation (CAT1), 31 participants (18\%) received an invitation accompanied by a cat picture (CAT2), and 70 participants (41\%) received a personalized invitation along with a cat picture (CAT1 and CAT2). The response rates per condition are presented in Table \ref{tab:responserates}. The descriptive statistics suggest that the response rates are higher before sending the reminders, indicating that motivated participants will carry out the task straight away. As earlier, participants who received multiple affective triggers (CAT1 and CAT2) had the highest response rate compared to the rest.

\begin{table*}[h]
\begin{center}
\begin{tabular}{L{2.5cm}lrrrlrr}
\toprule 
 & \multicolumn{1}{c}{\textbf{Control Condition}} & \multicolumn{3}{c}{\textbf{Experimental Condition}} & \multicolumn{3}{c}{\textbf{All Responses}}\\
    \addlinespace[0.1cm]
   & & \multicolumn{1}{c}{CAT1} & \multicolumn{1}{c}{CAT2} & \multicolumn{1}{c}{CAT1 and CAT2} \\
    Before Reminders        &  13 (1.48\%) & 28 (2.27\%) & 23 (2.64\%) & 39 (3.11\%) & 103 (2.43\%)    \\
    After Reminders &  10 (1.16\%) & 19 (1.57\%) & 8 (0.94\%) & 31 (2.55\%) & 68 (1.65\%)   \\
    Total Responses  &  23 (2.63\%) & 47 (3.81\%) & 31 (3.56\%) & 70 (5.58\%) & 171 (4.04\%) \\

    \bottomrule
    
\end{tabular}
\end{center}
\caption{Number of complete survey responses per condition before and after the reminder that was sent 14 days after the initial invitation. In parentheses, we provide the response rate per condition, that is, the percentage of responses per number of participants for each design group.}
\label{tab:responserates}
\end{table*}

\subsubsection{CAT Impact on Complete Participation}
The results of the regression analysis for complete participation (Table \ref{tab:partialregr}) suggest that invitations accompanied by a cat picture (CAT2) had a statistically significant impact on responses. The odds of responding to an invitation accompanied by a cat picture were 40\% higher than the odds of responding to an invitation without a cat picture.
Personalized invitations (CAT1) had a statistically significant impact on complete responses as well: the odds of responding to a personalized invitation were 46\% higher than the odds of responding to a non-personalized invitation.

We repeated the analysis for exploring the potential interaction between CAT1 and CAT2, and compared the two models (with and without the interaction term). As earlier, the two models were not significantly different, and therefore, we can ignore the interaction term $CAT1*CAT2$ (Table \ref{tab:interact}, and Appendix \ref{sec:app:interaction}, Table \ref{tab:anovamodels}).

\begin{table*}[h]
\begin{center}
\begin{tabular}{lcccccc}
\toprule 
\multicolumn{1}{c}{\textbf{Random Effects}}& \multicolumn{3}{c}{\textbf{Partial Participation}} & \multicolumn{3}{c}{\textbf{Complete Participation}} \\
    \addlinespace[0.1cm]
    & \multicolumn{1}{c}{Variance} & \multicolumn{1}{c}{Std. Dev} & \multicolumn{1}{c}{} & \multicolumn{1}{c}{Variance} & \multicolumn{1}{c}{Std. Dev} & \multicolumn{1}{c}{}\\
Reminder(Intercept) & 3.373  &1.836  & 3.003  &  1.733     \\

   \multicolumn{1}{c}{\textbf{Fixed Effects}}& \multicolumn{3}{c}{} & \multicolumn{3}{c}{} \\
    \addlinespace[0.1cm]
    & \multicolumn{1}{c}{Coeffs} & \multicolumn{1}{c}{Std. Error} & \multicolumn{1}{c}{z-value} & \multicolumn{1}{c}{Coeffs} & \multicolumn{1}{c}{Std. Error} & \multicolumn{1}{c}{z-value} \\
    
    Intercept  & -1.91  & 1.32 & -1.45 &-2.24 $^{.}$ & 1.25 & -1.8\\
    CAT1|1  & 0.54 $^{*}$ & 0.24 & 2.19 & 0.41 & 0.27 &  1.24 \\
    CAT2|1      & 0.31  & 0.27 & 1.16 &0.37  & 0.29& 1.48\\
    CAT1|1*CAT2|1 & 0.05 & 0.33 & 0.15 & -0.045  & 0.36& -0.12\\
  
    \bottomrule
    \multicolumn{5}{l}{\scriptsize{$^{***}p<0.001$, $^{**}p<0.01$, $^*p<0.05$, $^.p<0.1$}}
\end{tabular}
\end{center}
\caption{Regression models for partial participation and complete participation that explored the interaction of affective triggers CAT1*CAT2 regarding response behavior. Email reminders were modeled as random effects.}
\label{tab:interact}
\end{table*}

\subsubsection{Response Bias Due To Affective Triggers}
Our findings showed that participants who received CAT-enhanced invitations demonstrated higher response rates than those who did not. Nonetheless, affective triggers may also have an impact on survey responses in terms of ratings, thus inducing response bias~\cite{furnham1986response}. To this end, we studied and compared the responses of participants in the control and experimental conditions. In particular, we conducted a one-way between subjects ANOVA to compare the effect of affective triggers on survey responses.

We used Levene's test~\cite{levene1960contributions} to check the homogeneity of variances which confirmed that the ANOVA assumption for variance equality is satisfied ($F(3,167)=2.29, p=0.08 > 0.05$; the p-value is not less than the significance level of 0.05, meaning that we can assume the homogeneity of variances). To assess the assumption that the residuals are normally distributed, we used a normality (Q-Q) plot. The residuals followed approximately a straight reference line suggesting normal distribution (Appendix \ref{sec:app:qqplot}, Figure \ref{fig:qplot}).

The ANOVA results did not indicate statistically significant effects of the affective triggers on the average survey score per category and overall ($F(3,167)=0.439, p=0.726 > 0.05$). Thus, we cannot reject the null hypothesis. This suggests that we cannot conclude that the use of affective triggers has an impact on survey responses in this study by introducing undesirable bias.

\subsection{Validation results}
\label{sec:results:validation}
Here, we present the results of the validation survey. With this survey, we aimed to clarify the following: a) the impact of different triggers, such as personalized salutations and pictures, on survey participation intentions (C1), and; b) the choice of the animal picture as a means to promote participation (C2 and C3).

Out of 100 participants, 38 identified as female, 61 as male, 0 as non-binary, and 1 did not disclose their gender. Regarding participants' age, the majority of participants (37 participants) reported being between 35 and 44 years old, followed by fewer participants reporting to be between 25 and 34 (32 participants) and 45 and 54 (15 participants) years old. 10 participants reported being between 55 and 64, 5 participants between 65 and 74 and 1 participant between 18 and 24 years old. 

\subsubsection{Impact of survey design elements on participation intention (C1)}

The analysis of the participants' responses to the validation survey showed that the realistic time estimate as a survey design element received the highest rate in terms of participation intentions ($M=4.01$ on a 5-point Likert scale, $SD = 1.15$) while the attached picture was rated lowest ($M=2.95$ on a 5-point Likert scale, $SD = 1.42$). 
The regression analysis results showed that participation intentions were statistically significant and positively associated with design elements such as the targeted audience scarcity, realistic time estimate, and the attached cat picture. This shows that participants who indicated they would participate in the survey because the invitation included elements such as the cat picture (CAT2) or the realistic time estimate also demonstrated higher -- or stronger -- participation intentions. On the contrary, providing the source of information, personalized salutations, and the investigator's information did not have a statistically significant effect on participation intentions. This means that although personalized salutations (CAT1) and investigator's information, as well as the source of information were highly rated by participants, this was not reflected in the reported participation intentions.
The results are presented in Table \ref{tab:evalA}.

\begin{table*}[h]
\begin{center}
\begin{tabular}{lccc}
\toprule 
   \textbf{Survey Element} & \textbf{Average Rating (SD)} & \textbf{Coeff} & \textbf{Std. Error} \\
   Realistic time estimate  & 4.01 (1.15)& 0.653$^{**}$& 0.28\\
   Investigator's information & 3.81 (1.21)& -0.186& 0.23\\
   Targeted audience scarcity& 3.78 (1.22) & 0.524$^{**}$& 0.23\\
   Personalized salutation  & 3.45 (1.19)& 0.02& 0.21\\
   Source of information & 3.36 (1.35)& 0.311& 0.19\\
   Attached picture  & 2.95 (1.42)& 0.957$^{***}$& 0.2\\
   
    \bottomrule
    \multicolumn{4}{l}{\scriptsize{$^{***}p<0.001$, $^{**}p<0.01$, $^*p<0.05$, $^.p<0.1$}}
\end{tabular}
\end{center}
\caption{Average participation intentions on a 5-point Likert scale anchored between "strongly disagree" (1) and "strongly agree" (5) and regression coefficients and standard errors for participation intentions. The survey elements are sorted by their average rating, from highest to lowest.}
\label{tab:evalA}
\end{table*}

\subsubsection{Participation intentions, affect and pictures of small animals (C2, C3)}
Next, we wanted to clarify the relationship between individual preferences for pictures of cats, dogs, and small animals and the cat picture as a survey design element (C2). Moreover, we wanted to clarify the connection between the theme of the picture attached to the survey invitation (that is, the picture of the cat) and individual participation intentions (C3). To that end, we analyzed potential relationships between the preference of participants for pictures of cats, dogs, and small animals, and the rating of the cat picture as a design element for the survey invitation (OLR\_picture) and potential relationships between the preference of participants for pictures of cats, dogs, and small animals, and participation intentions (OLR\_participation).
The results -- presented in Table \ref{tab:evalB} -- confirmed that people who enjoy cat pictures in general rated higher the use of the cat picture as an attachment in the invitation email (C2). However, we did not find any statistically significant effect for people who prefer pictures of dogs or small animals. In other words, we cannot conclude whether people who enjoy pictures of dogs or other small animals would react positively or negatively to the attached cat picture.
In terms of participation intentions (C3), the results suggest that people who generally enjoy cat pictures demonstrate higher participation intentions. We did not find any indications for positive or negative impact on participation intentions among people who like pictures of dogs or pictures of other small animals.
These findings suggest that on the one hand, attaching a picture of a cat to an invitation email may indeed support participation intention. On the other hand, there was no indication that the use of a cat picture would negatively impact participation intentions or affect the way people react to the invitation design elements.
\begin{table*}[h]
\begin{center}
\begin{tabular}{lccc}
\toprule 
    \textbf{} & \textbf{User Rating Average (SD)} &  \textbf{Coef OLR\_picture} & \textbf{Coef OLR\_participation} \\
   Pictures of Cats & 7.52 (1.72) & 0.788$^{***}$ & 0.375$^{***}$\\
   Pictures of Dogs & 7.25 (1.84)& -0.012& 0.196\\
   Pictures of Small Animals & 7.13 (1.68)& 0.125& 0.269\\
   
    \bottomrule
    \multicolumn{4}{l}{\scriptsize{$^{***}p<0.001$, $^{**}p<0.01$, $^*p<0.05$, $^.p<0.1$}}
\end{tabular}
\end{center}
\caption{Average ratings for pictures of cats, dogs, and small animals on a scale from  "dislike extremely" (1) and "like extremely" (9) and regression coefficients among ratings of an attached picture as a survey invitation design element and participation intentions.}
\label{tab:evalB}
\end{table*}

\section{Discussion}
\label{sec:discussion}
In this section, we discuss our findings with respect to the research questions (section~\ref{sec:intro}) and hypotheses (section~\ref{sec:method}). The survey response rates in this work varied from 1\% to 9\%, thus being in the realm reported in prior research~\cite{koo2005challenges, zillmann2014survey}. 

\subsection{Do CATs have a significant impact on online survey response rates? (RQ1)}
Our findings indicate that personalized invitations and invitations accompanied by a cat picture had a positive impact on response rates. However, their impact varied depending on invitation strategies, such as the use of reminders, and with respect to response behavior (for example, accessing and partially responding vs. completing the survey). 
Addressing participants using their first name had a significant positive effect on partial and complete participation -- confirming hypothesis \rehyp{H1} -- although the effect appeared to be stronger for partial participation. The descriptives showed that participants who received personalized invitations responded to the survey after receiving the first invitation suggesting that the effect was time-sensitive and potentially short-lived: participants were eager to browse the survey directly upon receiving the invitation but this impulse faded out quickly, leading to incomplete responses. Our results add to the findings reported by Heerwegh~\cite{heerwegh2005effects}, who found personalized salutations to increase response rates while not influencing survey completion in the context of higher education. Moreover, they hint at personalization potentially fostering a parasocial interaction between the researcher and the participant. Parasocial interaction refers to a relationship experienced by members of an audience in their encounters with, for example, performers in mass media~\cite{horton1956mass}. The effect of personalization on parasocial interaction has also been observed outside of studies on mass media, particularly in the context of education~\cite{gray2017designing}. Thus finding could thus serve as a basis for future work.

Invitations accompanied by a cat picture had a statistically significant positive impact on response rates, which appeared to be constant between partial and complete responses. This may indicate that artifacts, such as a cat picture, can be a more robust drive over time to complete a survey than personalization. Anecdotally, we note that the surveyor received -- among comments and questions regarding the survey content -- one message referring to the surveyor's cat: \textit{``I must say this I really found funny and amusing about "Me and my cat are looking forward to your answers, and thank you for your time and help!" and the picture of your cat. It was a really big survey form and I filled-up as much as possible."} We report this message as an example that demonstrates our working hypothesis about how CATs may promote feelings of personal connection between the participants and the surveyor~\cite{oliver2011entertainment} while taking into account the low response rates of the survey and the spontaneous nature of this action. We retrieved no other comments (neither positive nor negative) regarding the use of the cat picture. Thus, we were able to confirm hypothesis \rehyp{H2} and add to existing work that suggests abstaining from sending invitations that contain graphical content~\cite{kaczmirek2005web}. 

We saw that the impact of personalized invitations on response rates might decrease over time and in terms of participation mode (partial vs. complete) while the effect of invitations accompanied by a cat image appears to be constant. We attribute this finding to two related aspects:
\begin{enumerate}
    \item The novelty effect: Personalized invitations are quite common nowadays. This is not the case for the cat-image invitation. Thus, the effect of the personalized invitation may be short-term and easier to wear out over time due to being “not-so-novel” compared to the image invitation;
    \item The picture superiority effect: The effect of pictures vs. words may extend from coding and retrieval of short-term memory~\cite{curran2011picture} to affect and motivation, potentially suggesting that pictures have a greater and longer-lasting impact than words on affect and cognition.
\end{enumerate}
 
The validation survey partly confirmed our findings. Compared to survey design elements that do not target participants' affect, such as providing a realistic time estimate or providing the researcher's contact information~\cite{heerwegh2005effects,crawford2001web}, participants rated the CAT objects lower in terms of attributing their participation intention to these elements. That is, they stated they were not particularly inclined to participate \textit{because} of the specific elements. However, an analysis of participation intentions and design elements ratings showed that the attached picture had a positive and statistically significant impact on participation intentions. This may suggest that CAT triggers influence intentions without participants consciously realizing or controlling. Furthermore, we did not find any evidence indicating that the theme of the attached picture may have a negative effect on participation intentions.

We focused on limited examples of CATs, both in terms of themes (i.e. cats) and media types (i.e. pictures). We assume that CATs, as a concept, will probably also work for themes that can trigger positive emotions -- such as idyllic landscapes depicting a sunset or pictures of decorated gourmet dishes -- as long as they are generic and not directly or implicitly related to the target of the survey \cite{kaczmirek2005web}. One may expect that the media type of a CAT could influence its impact due to the media type's inherent or external attributes -- for example, how it is cognitively processed or whether it is directly visible or needs further action in order to be viewed (that is, a video or audio file that would require to be downloaded and then played).

\subsection{When combined, do CATs have an additive effect on online survey response rates? (RQ2)}
A subgroup of participants received both kinds of triggers. In this case, our analysis did not reveal a statistically significant interaction between CATs that would suggest potential additive effects on response rates. Thus, \rehyp{H3} was not confirmed. 

Although one could assume that personal salutations and accompanying pictures can act complementary in a setting like the one we studied and therefore, their combined use can strengthen their positive effect, other interpretations are plausible. For example, the cat picture affecting participation due to its hedonic tone (a cat being cute or fun) or due to its novelty rather than due to promoting a personal connection between the participants and the surveyor. These different effects could thus potentially explain why we found no evidence to indicate an interaction (either denoting a synergistic or multiplicative effect) between CAT1 and CAT2.

\subsection{Do CATs introduce bias on online survey responses? (RQ3)}
Our analysis suggested that the survey responses within groups and conditions did not differ significantly either per study category or overall. The participants' responses were on average similar and did not demonstrate significant variation regarding whether they had received a CAT-enhanced invitation or not. This indicates that CATs did not induce any response bias and that \rehyp{H4} cannot be confirmed. CATs thus do not appear to suffer from the same issue as e.g. monetary incentives that can induce response bias~\cite{mizes1984incentives}.

\subsection{Theoretical and Practical Implications}

This work has a number of implications for research and practice. Our findings can serve as hints and practical guidance for online survey designers who are interested in improving survey response rates. Survey designers can expand the use of affective triggers in order to simulate an interpersonal link between participants and surveyors. 
Furthermore, this work can be used as a basis to extend the use of affective triggers to online participatory environments, for example crowdsourcing platforms, in order to foster participation, commitment, and consistency.

Regarding generalisation of this work, Myrick~\cite{myrick2015emotion} demonstrates that people who enjoy viewing cat-related media also enjoy watching videos or viewing pictures of dogs or other (small) animals. Thus, we expect our findings to hold for such pictures in general. 

Further studies, both in relation to the design of affective triggers as well as incorporating experiments are needed in order to provide evidence about the relationship between affect and response rates. In order to guide research towards this direction, questions such as \textit{``What other aspects of affect, such as joy or contentment, could we facilitate or promote in order to influence participation?"} or \textit{"What is the effect of the long-term use of affective triggers on participation?"} could be further explored.

\subsection{Limitations}
Our work was based on online surveys that studied events with a local character. This means that the majority of participants can be expected to share a common cultural background and characteristics, which in turn might have an impact on our findings. Affect -- and in particular affective triggers like the ones we used -- may have a different connotation or impact depending on participants' cultural backgrounds. Thus we cannot exclude the possibility that the specific CATs used here had the described impact due to the cultural characteristics of our participants. In order to address this limitation, further studies with an international character are needed.

With respect to CATs design, we used two very specific approaches. On one occasion, we addressed the participants using their first name. This method is very common, especially for targeting individuals in marketing and business contexts. Alternatively, we used the picture of the surveyor's cat along with a related message. One could argue that this is a very limited sample of CATs, which in some cases might not even trigger participants' affect in relation to a specific design aspect. To address this limitation, we consider designing a pool of CATs and evaluating their impact on various aspects of affect before using them as survey design features as well as to control for confounding factors in future experiments.

Regarding the invitation design, we followed established recommendations to maximize participation. We chose design elements based on their simplicity and ease of implementation. However, more sophisticated approaches could be adopted to design invitations. Specifically, when it comes to introducing CAT1 and CAT2, one may argue that the email client used to read the invitation could have affected the result: for example, the personal salutation could have been visible without viewing the email invitation while the cat picture would be hidden unless the receiver would view the email since it was sent as an attachment. We acknowledge this limitation and the need for further studies to explore how design aspects, like the ones mentioned above, could affect participation.

Concerning population sampling, we randomly assigned participants to the different experimental groups (conditions) to avoid potential bias since the affective state of our target population was unknown. Our study was conducted in the context of online hackathons, which implies certain age groups, gender distribution, cultural and educational backgrounds. Our population is thus not representative of the general population. To address this limitation, further studies with representative samples are necessary.

Regarding the validation survey, we provided participants with an artificial scenario. The artificial scenario in combination with the nature of the experimental context (crowdsourcing platform with low monetary rewards for participants) may have impacted the relevance or validity of the findings. However, since our aim was not to compare the findings of the main and the validation survey but rather to clarify certain aspects, the second survey can be perceived as self-contained. Moreover, we asked direct questions about participants' intentions. This forces participants to reflect on how they would react in a certain situation rather than observing them which would allow us to draw insights regarding their behavior. Observation is not feasible in this setting since it would not be possible for us to observe which parts of the invitation would have a stronger effect on participants' intentions, and at the same time our aim was to compare between items that were assessed in the same way. We also relied on single items, which can be problematic with respect to reliability.

Finally, we acknowledge that this work requires further investigation in order to map and explain the relationship between casual objects, affect, and human reaction. At the same time, we acknowledge the need for additional measures to prove the potential relationship between positive emotions and survey participation and measures regarding the affective state of participants upon receiving the invitation. However, we want to note that we considered it important to explore the use of CATs on response rates without introducing survey items unrelated to the survey topic. We perceive this appropriate to not draw additional attention to the CAT itself and introduce bias. However, we envision that this work can provide hints to survey designers and researchers both with respect to user behavior and practice and online participation. 

\section{Conclusion}
\label{sec:conclusion}
Research on online surveys response rates focuses on improving response rates by targeting participants through personalization, pleading for their support, or promoting and capitalizing on the feeling of belonging to a community. Prior studies have explored the role of survey design in terms of items, aesthetics, or both. Here, our objective was to explore the role of affect in this context. In particular, whether and how we can improve online surveys response rates by using simple objects that target participants' affect, or in other words, casual affective triggers (CATs) in invitations to online surveys. To that end, we carried out a study where we used two different kinds of CATs individually as well as in combination. Our findings suggest that CATs can impact online survey response rates.

Invitations accompanied by a cat picture (CAT2) had a significant impact on complete participation, meaning that participants who received CAT2 completed their surveys more frequently compared to other participants and regardless of reminders. CATs that target personalization (CAT1) generally appeared to significantly impact partial and complete participation. The effect was stronger for partial participation though. Lastly, the combined use of CATs did not appear to have an additive effect on response rates.

Our findings were partly validated by a follow-up survey. In particular, we found that CATs -- and specifically the attached picture -- can have a positive impact on survey participation intentions. On the other hand, the theme of the attached picture did not seem to have a negative influence on participation intentions.

\begin{acks}
This research was partially funded by the Estonian Research Council (PRG1226).
\end{acks}

\bibliographystyle{ACM-Reference-Format}
\bibliography{references}
\newpage
\appendix
\section{Appendix}
\label{sec:app}

\subsection{Validation survey instrument}
\label{sec:app:instrument}

\begin{table}[h]
\begin{center}
\begin{tabular}{|L{\linewidth}|}
\hline
\rowcolor{Gray}
Intention to participate, anchored between strongly disagree (1) and strongly agree (5) \\
\hline \\ [-2.2ex]
I would participate in the survey if I would receive an invitation that looks like this. \\
[0.8ex]
\hline
\rowcolor{Gray}
Perception of triggers, anchored between strongly disagree (1) and strongly agree (5) \\
\hline \\ [-2.2ex]
I would participate in this survey mainly because... \\
\tableIndent...it is only directed at the participants of this event.\\
\tableIndent...it explains where the surveyors obtained the information to contact me.\\
\tableIndent...it provides a realistic time estimate.\\
\tableIndent...it provides contact information of the investigator.\\
\tableIndent...if you read this please choose strongly disagree for this option.\\
\tableIndent...it is personalized.\\	\tableIndent...of the attached picture. \\
[0.8ex]
\hline
\rowcolor{Gray}
Perception of animal pictures, anchored between dislike extremely (1) and like extremely (9) (5)~\cite{peryam1957hedonic}\\
\hline \\ [-2.2ex]
Overall how much do you like or dislike... \\
\tableIndent...pictures of cats\\			
\tableIndent...pictures of dogs\\
\tableIndent...pictures of small animals\\
[0.8ex]
\hline
\rowcolor{Gray}
Demographics \\
\hline \\ [-2.2ex]
How old are you currently? (18 to 24, 25 to 34, 35 to 44, 45 to 54, 55 to 64, 65 to 74, 75 or older, Prefer not to say) \\
Are you...? (Female, Male, Non-binary, Prefer not to say) \\
Do you consider yourself a minority? (For example in terms of race, gender, expertise or in another way) (Yes, No, Prefer not to say) \\ [0.8ex]
\hline
\end{tabular}
\end{center}
\caption{Scales utilized for the validation survey}
\label{tab:app:instrument}
\vspace{-10pt}
\end{table}

\subsection{Interaction analysis for CATs}
\label{sec:app:interaction}

\begin{table}[h]
\begin{center}
\begin{tabular}{lccccccc}
\toprule 
\multicolumn{1}{c}{\textbf{Models}} & \multicolumn{7}{l}{Without Interaction: Response ~ CAT1 + CAT2 + (1 | reminder)} \\
\multicolumn{1}{c}{} & \multicolumn{7}{l}{With Interaction: Response ~ CAT1 + CAT2 + CAT1*CAT2 + (1 | reminder)} \\

    \addlinespace[0.1cm]
    \multicolumn{8}{c}{\textbf{Partial Participation}}\\
     \multicolumn{1}{c}{} & \multicolumn{1}{c}{npar} & \multicolumn{1}{c}{AIC} & \multicolumn{1}{c}{BIC} & \multicolumn{1}{c}{logLik} & \multicolumn{1}{c}{deviance}
    & \multicolumn{1}{c}{Chisq} & \multicolumn{1}{c}{Pr(>Chisq)} \\
Without Interaction  & 4  &  1508.1 & 1533.5 &  -750.03  & 1500.1  &  &  \\
With Interaction  & 5  &  1510.0 & 1541.8 &  -750.02  & 1500.0  &  0.0247 & 0.8752 \\

\multicolumn{8}{c}{\textbf{Complete Participation}}\\
    \addlinespace[0.1cm]
     \multicolumn{1}{c}{} & \multicolumn{1}{c}{npar} & \multicolumn{1}{c}{AIC} & \multicolumn{1}{c}{BIC} & \multicolumn{1}{c}{logLik} & \multicolumn{1}{c}{deviance}
    & \multicolumn{1}{c}{Chisq} & \multicolumn{1}{c}{Pr(>Chisq)} \\
Without Interaction  & 4  &  1228.1 & 1253.5 &  -610.05  & 1220.1  &  &  \\
With Interaction  & 5  &  1230.1 & 1261.8 &  -610.04  & 1220.1  &  0.0156 & 0.9006 \\

    \bottomrule
    \multicolumn{5}{l}{\scriptsize{$^{***}p<0.001$, $^{**}p<0.01$, $^*p<0.05$, $^.p<0.1$}}
\end{tabular}
\end{center}
\caption{ANOVA comparisons for the regression models with and without interaction terms for partial participation and complete participation.}
\label{tab:anovamodels}
\end{table}

\newpage
\subsection{Q-Q Normality plot of residuals}
\label{sec:app:qqplot}
\begin{figure}[htb]
\centerline{\includegraphics[width=\linewidth,keepaspectratio,]{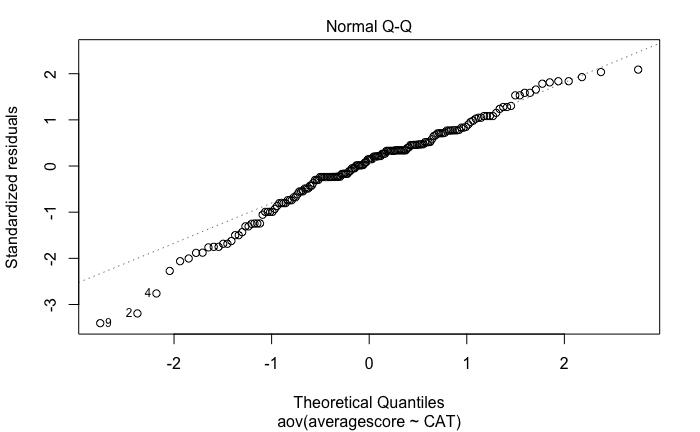}}
\caption{Normality plot of residuals: the quantiles of the residuals are plotted against the quantiles of the normal distribution along with a 45-degree reference line.}
\label{fig:qplot}
\Description{The Q-Q Normality plot that shows the plotted points non-decreasing when viewed from left to right and following the 45-degree line which is also plotted for reference. Some outliers are visible at the two ends of the plotted sequence.}
\end{figure}

\end{document}